\documentstyle[12pt,epsf,aaspp4]{article}

\tightenlines

\newcommand{\etal}{et~al.\ }

\begin{document}

\title{TESTING COSMOLOGICAL MODELS BY GRAVITATIONAL LENSING: \break
		I. Method and First Applications}
\author{Joachim Wambsganss}
\affil{Astrophysikalisches Institut Potsdam, 
	14482 Potsdam, Germany} 
\affil{ jwambsganss@aip.de }
\author{Renyue Cen and Jeremiah P. Ostriker}
\affil{Princeton University Observatory, Princeton, NJ 08544 }
\affil{ cen, jpo@astro.princeton.edu}

\begin{abstract} 

Gravitational lensing directly measures mass
density fluctuations along the
lines of sight to very distant objects. 
No assumptions need to be made concerning bias, the
ratio of fluctuations in galaxy density to mass density.
Hence, lensing is a very useful tool to study
the universe at low to moderate redshifts.

We describe in detail a new method to trace light rays 
from redshift zero through
three dimensional mass distribution to high redshift.
As an example, this method is applied here to a standard cold dark matter
universe. We obtain a variety of results, some of them statistical in
nature, others from rather detailed case studies of individual ``lines
of sight".  Among the former are the frequency of multiply imaged
quasars, the distribution of separation of the multiple quasars, and
the redshift distribution of lenses: all that as a function of quasar
redshift. We find effects from very weak lensing to highly magnified
multiple images of high redshift objects, 
which, for extended background sources,
(i.e. galaxies), range from slight deformations of the shapes
through tangentially aligned arclets up
to giant luminous arcs.

Different cosmological models
differ, increasingly with redshift,
in their predictions for the mass (thus gravitational potential)
distributions.
Our ultimate goal is to apply this method to a number of
cosmogonic models and to eliminate some models whose 
gravitational lensing properties are inconsistent with those observed.

\end{abstract}
\keywords{gravitational lensing, dark matter, cosmology,
	methods: numerical}

\bigskip
\bigskip
%
%
%
%
\section{Introduction} 
Each specific model for the development of cosmogonic structure
(e.g. the hot dark matter [HDM] or cold dark matter [CDM] scenarios)
has one free parameter, the amplitude of the density
power spectrum.
In the light of the COBE observations
(Smoot \etal 1992; Gorski \etal 1994),
it occurs, for the first time,
that this parameter is fixed ($\pm 12\%$) by the
determination
on the $5^o-10^o$ scale in the linear regime.
With its amplitude fixed, a secure determination
of the potential fluctuation on any other scale provides a test for
a particular cosmological model. 
Any single conflict between the theory
and reality can falsify the former.
The most leverage is obtained for tests which are made on 
angular scales as far away as possible from the COBE measurements.
But they should not be so small as to be greatly influenced 
by the difficulty inherent in modeling the physics of the gaseous,
baryonic components ($\le 10kpc$).
Thus, critical tests are best made on scales $0.01 Mpc<r<1Mpc$.
Gravitational lensing measures {\it directly} the
fluctuations in the gravitational potential along random lines
of sight to distant objects.
In contrast,
the conventional tools for comparing cosmogonic theories
with observations rely on either galaxy
density or galaxy velocity information, both of which
unavoidably suffer from the uncertain situations with regard
to density or velocity bias of galaxies over
the underlying mass distribution,
hampering our attempts to understand the more ``fundamental"
question concerning the mass evolution and distribution.
Hence gravitational lensing provides a powerful,
independent test of cosmogonic models,
because of its unique ability to directly
measure the gravitational potential fluctuations. 
The sampled properties of cosmic structure on certain scales
are also more ``fair" than other conventional measures owing to
the fact that lines of sight to distant objects are {\it random}
with regard to the foreground matter distribution.

We present here  a method to determine 
gravitational lensing effects for realistic mass distributions
from the weak to the strong lensing regime.
First, we fill the universe ``densely" with
adjacent mass cubes, taken from cosmological simulations,
along the line of sight,
where inside each cube the matter is
projected onto the middle plane. 
Then,
we study the lensing properties of the intervening matter
by ``shooting" light rays through the lens (mass) 
planes representing matter
distributions along redshifts (cf. Figure 1
for a schematic version of our actual scheme
[which uses of order 100 lens planes]).
The gravitational lens effects of  many lens planes 
have been studied by several authors  in the past (e.g.,
Kochanek \& Apostolakis  1988;
Schneider \& Weiss 1988a,b;
Jaroszy\'nski 1989, 1991; 
Jaroszy\'nski \etal 1990;
Lee \& Paczy\'nski  1990;
Babul \& Lee 1991;
Bartelmann \& Schneider 1991).
These authors have typically used point lenses or isothermal spheres
which were randomly distributed on individual lens planes.
The number of lens planes used
ranges from 2 to 32. 
Most of these studies involved the regime of weak lensing, i.e. slight
effects on the magnification of sources, but not multiple imaging.
Rauch (1991) studied the microlensing effect of three-dimensionally
distributed point lenses on high redshift supernovae.
A complementary approach to the one described here,
a semi-analytical   method using the 
Press-Schechter formalism was first used by Narayan and White (1987),
and more recently applied by Kochanek (1995) to various cosmological 
models.

A very preliminary attack on the problem
of studying the lensing properties of different cosmological models
has been presented by Cen \etal (1994, hereafter CGOT).
In that work no ray tracing was done.
Rather it was simply checked whether
or not mass accumulations were greater than the
critical level (cf. Turner, Ostriker \& Gott 1984, henceforth TOG)
at which multiple imaging will occur.
The mathematical and physical treatment here
is far superior to that adopted
in CGOT; the essential conclusions for the standard
CDM scenario, however,  are unchanged.
Some of these conclusions have already been presented
in Wambsganss \etal \ (1995).
Here we provide the mathematical basis of that method
as well as some tests of the method.

We use cosmological simulations for the standard  COBE-normalized
$\Omega = 1$, CDM universe 
as an example to illustrate the method. 
A first report on the results has been published elsewhere
(Wambsganss \etal\ 1995). 
In principle, mass distributions from any
cosmological model can be treated with this method.
The ultimate goal of this project is to provide
a quantitative comparison of the lensing properties
of different cosmological models 
(e.g., CDM with $\Omega = 1$; low density (open) CDM models;
mixed dark matter [MDM] models; flat models with a cosmological 
constant; isocurvature models etc) 
with observations.
For example, we anticipate that different cosmological
models, which are typically tuned to
give approximately the right properties
when being compared with our local universe at redshift $z=0$,
will show different lensing properties because
they reach the final state ($z=0$)
from quite different paths.
Thus, the typical lens will appear at different redshifts in different
models. For example, in MDM models, where structure forms late,
the lenses will have small redshift, but in isocurvature models,
where the first structures form shortly after recombination,
the average lens redshift will be much larger.

For many questions (e.g. the fraction of multiply imaged quasars as 
a function of angular separation, magnification ratio, redshift) 
one needs a large number of simulations with different realizations
of the matter distribution, in order to give quantitative answers
on a good statistical basis. For other applications, e.g. detailed
studies of lensing by
clusters of galaxies and the effects of chance  alignments
with foreground or background matter, 
a specific realization must
be studied in detail.

While it is in principle possible to 
simulate, numerically in one simulation,
the entire path from observer to source
(very large simulations with elongated geometry have
been done; e.g., Park \& Gott 1991),
technical limitations preclude this at the present time.
For now we perform many independent simulations
for different volumes along the line
of sight (since repeated structures are obviously
to be avoided).
We have developed an approximate technique for this which we term
the ``convolution method".
Physically it consists of assuming that separate pieces of an inhomogeneous
universe act as if they were part of an homogeneous
universe of the same mean density as themselves.
Details of the method, which is an improved version over that
applied in CGOT, are presented in the Appendix,
as well as detailed checks which indicate the accuracy
of the approximations involved.

Although we think this convolution technique describes the properties
accurately enough on the scales in which we are interested,
we hope in the future to be able to avoid this step and be able to
use simulations that have a large enough dynamic range  to cover
a representative part of the universe and have high enough resolution
in order to represent the potential wells of galaxies, groups and 
clusters appropriately.
The analytical techniques presented in this paper
for the determination of the lens properties
are independent of the details
of the method for computing the evolution.
We will use the same ray tracing method when later we utilize higher
resolution numerical evolution techniques which do not
depend on the convolution method.

The rest of the paper is organized as follows.
The ray-shooting technique is described and discussed
in detail in Section 2. 
As a result of this ray-shooting,
we obtain a mapping of the rays from the 
image plane (sky plane) to the 
source plane.
With this mapping it is possible to 
determine the magnification distribution of
background sources, the fraction of the source
plane that is multiply imaged,  the positions and shapes as well
as the topology
of the caustics and the corresponding critical lines (Section 3). 
For a set of given source positions and shapes 
in the source plane at a given redshift, we determine and analyze
the corresponding image configurations as they would appear in the
sky, as is  shown in Sections 4 and 5.
Finally in Section 6 we give a short summary
and describe the planned 
applications and the types of predictions we will be able to 
make for the different cosmological models.

%
%
%
\section{Multiplane Lens Equation, Ray Shooting, Tests}

In Schneider, Ehlers and Falco (1992, subsequently SEF) a whole chapter
deals with the theory of multiple light deflection,
i.e. gravitational lensing by more than one plane. 
We refer the interested reader to this text book for details. 
Here we just describe the multiplane lens equation as we use it:

$$
		\left( \begin{array}{c}
			y_{1}  \\
			y_{2}
			\end{array}   
		\right) = 
{\vec y} = {\vec x} - \sum_{i=1}^{N} {D_{is} \over D_s}
	{\vec \alpha}_i  ({\vec x})=  
		\left[ \begin{array}{cc}
			a_{11}({\vec x}) & a_{21}({\vec x}) \\
			a_{12}({\vec x}) & a_{22}({\vec x})
			\end{array}   
		\right]
		\left( \begin{array}{c}
			x_{1}  \\
			x_{2}
			\end{array}   
		\right) = 
		A({\vec x}) {\vec x}. 
		\eqno(1)
$$

\noindent Here 
${\vec x}$  is the position vector of a light ray in the
	first lens plane (or equivalently in the image/sky plane);
${\vec y}$  is its corresponding position in the 
source plane;
${\vec \alpha}_i$  is the (two dimensional) 
deflection angle  in lens plane
\# i; $D_s$ and $D_{is}$ are the angular diameter distances 
between observer and source plane and between 
$i$th lens plane and source plane, respectively.
The lens planes are labelled from $1$ through $N$ in order of 
increasing redshift.

The part of the equation that we are essentially interested in
is \  \ \ ${\vec y} = A({\vec x}) {\vec x}$\ \ \ ;
it describes just the mapping between
the image/sky plane and the source plane. The relation between the 
positions $\vec x$ in the image/sky plane and
$\vec y$ in the source plane is given by \ $A({\vec x})$, 
the Jacobian  matrix of the mapping. 
In the case of multiple lens planes the Jacobian 
$$
A({\vec x}) \equiv {\partial {\vec y} \over \partial {\vec x}}, 
\hskip 2.0truecm
A_{ij} \equiv {\partial y_i  \over \partial x_j}, 
\eqno(2)
$$
does not have all the nice
properties it has in the single lens plane case, in particular it
is not symmetric any more in the general case: $a_{12} \ne a_{21}$.
Furthermore it is not curl-free: ${\rm curl} A \ne 0$.
Many other properties, however, are still valid. The magnification
of a certain image at position $\vec x$ is  given as
$$
\mu(\vec x)      = \left[ \ \det A({\vec x}) \ \right] ^{-1}
	= \left[ \
	  a_{11}({\vec x}) a_{22}({\vec x}) 
	- a_{12}({\vec x}) a_{21}({\vec x}) \ \right] ^{-1}.
\eqno(3)
$$

We will see in later sections that other combinations of
the components of \ $\det A({\vec x})$ \ represent useful
quantities as well.
The effective surface mass density $\kappa$ (which is
basically the sum of the physical surface mass densities of
the individual lens planes, weighted by the
corresponding ratios between the angular diameter distances)
for any position ${\vec x}$ in the sky plane is given as
$$
\kappa(\vec x) = 1 - 0.5 \  [\ a_{11}({\vec x}) + a_{22}({\vec x}) \ ].
\eqno(4)
$$
\noindent
Similarly, the components of the shear 
$\vec{\gamma}({\vec x}) = [\gamma_1({\vec x}),\gamma_2({\vec x})]$
are 
$$
\gamma_1(\vec x) =  - 0.5 \ [\ a_{11}({\vec x}) - a_{22}({\vec x}) \ ];
\eqno(5a)
$$
$$
\gamma_2(\vec x) =  - 0.5 \ [\ a_{12}({\vec x}) + a_{21}({\vec x}) \ ].
\eqno(5b)
$$

In order to ``shoot" a light ray $i$ through a single lens plane, 
what must 
be done is to determine the deflection angle 
$\ \vec{\alpha_i} \ $ of this ray
due to all matter in this lens plane. 
For a number of $n$ point lenses this is just a summation of the 
deflection angle by each point lens $j$:
$$
\vec{\alpha_i} =  \sum_{j=1}^{n}\vec{\alpha_{ji}} = 
    {4 G \over c^2} \sum_{j=1}^{n} M_j { {\vec r_{ij}} \over r_{ij}^2}.
\eqno(6)
$$
Here $G$ and $c$ are the gravitational constant and the speed of 
light, respectively,  $M_j$ is the mass of point lens $j$, 
$\vec{r_{ij}}$ is the projected (vector) distance between the 
positions of
light ray $i$ and point lens $j$, and $r_{ij}$ is its absolute value,
$r_{ij} = \sqrt{(x_i - x_j)^2 + (y_i-y_j)^2};$
here ($x_i$,$y_i$) is the position of ray $i$, and
      ($x_j$,$y_j$) is the position of lens $j$.

The straightforward determination of one deflection angle takes
about ten floating point operations.
As a lens plane, we consider a two-dimensional distribution of matter 
that is given in a grid; for the examples here we will always use a 
regular matter grid with 500$^2$ positions
(in subsequent work we will increase the resolution to 
$800^2$ positions; in principle this can be increased to any desired
resolution, depending on the desired purpose).
This is the projection
of the three-dimensional matter distribution inside a cube with
comoving side length \  $L = 5 h^{-1} Mpc$ \ 
(where $h = H_0/100$ is the Hubble constant divided by 100)
onto one of the three faces of the cube.
We treat this matter distribution as 250,000 point lenses, or
more accurately, as ``smeared out" in matter or lens pixels, i.e.
in a square region of comoving $10 h^{-1} kpc$ at a side.
We must determine the deflection angle at least at as many
positions as there are lens pixels, i.e. at least on a 500$^2$ grid
(again, this is the resolution we use for the examples here;
this can be changed to whatever ray density is appropriate to the
problem at hand).
Because we will use of order 100 lens planes,
the total number of mathematical operations for a direct
determination of the deflection angle amounts to about 
$10 \times 500^2 \times 500^2 \times 100 
\approx O(10^{14})$ floating point operations. 
This is non-trivial, even for fast computers.

In order to speed up the calculation
we determine the deflection angle in each lens plane with 
a hierarchical tree code in two dimensions,
as described in Wambsganss (1990). The idea is to group together
lenses that are far away from the light ray, and treat these
``cells" (whose sizes increase with  the distance to the
ray) as pseudo-lenses. 
Nearby lenses are treated individually, lenses at intermediate 
distances to the light ray are grouped into small cells of a few lenses,
and those lenses very far away from the light ray considered, 
are in cells containing up to a few thousand lenses.
This tree code treatment speeds up the determination of the
deflection angle by a large factor. The number of lenses and
pseudo-lenses necessary for an accurate
determination of the deflection
angle in one plane is of order 100 
(rather than 250,000 for the direct determination).
There is some overhead, namely the determination of the tree structure
for each ray,
i.e. the configuration of which lenses have to be treated directly and
which ones are to be put in cells of different sizes. However, this
is done just once, and used again in all subsequent planes, since
the relative positions of the lenses are exactly the same.
The cost of this additional calculation is 
negligible compared to that of a direct determination 
of the deflection angles.

In its simplest application the tree code assumes that all matter
inside a cell is located at its center of mass. We do,   however,
use higher order multipoles of the mass distribution in 
order to increase the accuracy (cf. Wambsganss 1990). 
For the actual calculations
we use multipoles up to order 6. 
In  Figure 2 we show the deviation of the 
deflection angle determined with this tree-code method 
from the directly determined, i.e. accurate deflection angle:
$(\alpha_{direct} - \alpha_{tree})$ as a function of $\alpha_{direct}$ 
in arbitrary units (the top row indicates the x-component, the
bottom row the y-component).
The panels correspond to determinations of $\alpha_{tree}$
including multipoles of different order; they
represent: all matter in center of mass, i.e. only monopole term 
(leftmost column),
monopole term plus quadrupole moments (2nd column), 
monopole term  plus all multipoles up to order 4 (3rd column), and
monopole term  plus all multipoles up to order 6 (rightmost column).  
In the actual calculation of the deflection angle we use multipoles
up to order 6, corresponding to the rightmost panel.
It can be seen that for this case the largest deviation
between the two methods of determining the deflection angle
is of order $2 \times 10^{-3}$,  with the rms deviation quite
a bit smaller.  This seems to be a good enough approximation
for our purposes.

For point lenses the deflection angle formally diverges 
when the distance between light ray and lens position 
becomes zero, $r_{ij} \rightarrow 0$. 
However, in
our case this is an artificial divergence, because the underlying
mass distribution is smooth, we only approximate it by a large number
of point lenses. In order to avoid this artificial divergence,
we always determine the deflection angle at the points directly in the
center between four lens positions
so that nearest neighbor effects cancel. 
That means, for each lens plane we determine the deflection angles
for a regular grid of ``test-rays".
But, since
the real positions  of the light rays are offset from the positions 
of these ``test-rays", for which we determined the deflection angles,
we calculate  the 
actual deflection angles of the rays that are followed through the
planes using a bicubic 
interpolation between the four test-rays surrounding the real ray
position.
In this way we obtain a smooth two-dimensional field of deflection 
angles for each lens plane, as is to be expected for a continuous 
matter distribution.

The angular size of the (square) bundle of light rays 
that we consider is determined by the angular size of the highest
redshift lens plane that we use. As an example, the
angular size of a cube with comoving size $L = 5 h^{-1} Mpc$
at a redshift $z = 3$ is about $\beta \approx 350$ arcsec for
a standard CDM universe. 
The (average) angular size of the field of rays is fixed, 
it is the same for all lens planes; however,
as the physical size of the underlying cosmological cubes
is expanding with decreasing redshift, 
the angular size
of the lens planes increases rapidly with decreasing redshift.
This means that the field of real-rays intercepts only a small
part of the lens plane for small redshifts (cf. Figure 1).
This effect is reflected in the matter distribution inside the
field of view: for low redshift lens planes, one lens-pixel
(which has fixed comoving size) covers quite a large angular region.

The matter distribution of each lens plane inside the field of rays
is shown in Figure 3a for a particular line of sight:  
dark (light) gray means high (low)
surface mass density (in grams per cm$^2$).  From 
top left to bottom right you see the matter inside the beam
for 60 different lens redshifts, starting with $z \approx 0$
and increasing to $z \approx 3.0$. For the low redshift planes
one can easily identify the pixel size of the matter
distribution, because there one matter pixel (physical size
5 h$^{-1}$ Mpc $ /(1+z)/500 = 10 h^{-1}(1+z)^{-1}$kpc)
corresponds to many tens of arcsecs.
In Figure 3b the integrated matter distribution up to a
redshift $z \approx 3.0$ is shown. This is the ``physical" sum,
i.e. in units of mass per physical area. This map could be converted
into a map of light, if one assumes, e.g., a constant
mass-to-light ratio and  weights the individual lens planes
by the inverse of the (luminosity) distance.
If one compares Figures 3a and 3b, 
it is quite obvious that the total surface mass density is 
dominated by a few planes along the line of sight.
In general, lensing is a very convergent 
process:
it is not due to the sum of a large number of weak
lensing events;
rather it is due to the occasional passage of a light
ray past a large mass concentration.
This fact makes the requirements for statistical validation
of our results especially stringent.
Many independent realizations of the universe must
be made and many lines of sight examined before
one can have confidence (by examination of convergence) 
that the derived statistical results are robust
(i.e.  the statistically averaged results become independent
of the number of realizations studied).

For each lens plane the critical surface mass density $\Sigma_{crit}$
necessary
for multiple imaging can be calculated (cf. SEF):
$$
\Sigma_{crit} = {c^2 \over {4 \pi G } } { {D_s \over {D_i  D_{is}} } }.
\eqno(7)
$$
Comparing the surface mass
density in each matter pixel $(m,n)$ with this number identifies the
regions that should produce multiple images of a background source:
$$
\kappa_{m,n}  = \Sigma_{m,n} / \Sigma_{crit} =  
\Sigma_{m,n} * {4 \pi G \over c^2 } { {D_i  D_{is}} \over D_s},
$$
where $\Sigma_{m,n}$ is the  physical surface mass density in pixel
$(m,n)$, and $\kappa_{m,n}$ is its  normalized surface mass density.
When $\kappa_{m,n} > 1$, then multiple lensing occurs.
We note, however, that
overdensity in a single plane is a sufficient but not necessary
condition of multiple imaging; chance alignments of two 
or more individually subcritical but collective supercritical 
regions can produce multiple images as well as
slightly underdense regions in combination with a sufficiently
large value of shear. 
A comparison of such regions with the actual location of multiple images
serves as a good check of the whole method.

For each complete run of the light rays through all the lens planes,
the positions of the light rays in the image/sky plane and the 
corresponding positions in the source plane are stored. 
In subsequent sections we will describe
how the matter distribution is obtained
in each lens plane, and
how we analyze the mapping of light rays through essentially 
a realistic 3-dimensional matter distribution.
But first we like to add a few paragraphs discussing
the issue of numerical resolution, and describing
an additional test of our method.

Many authors have studied magnification cross sections
and caustic structures of various lens models. 
To  name just a few: 
Blandford \& Kochanek (1987) 
and
Kochanek \& Blandford (1987) 
studied magnification cross sections and probability distributions 
for isolated elliptical potentials;
Hinshaw \& Krauss (1987) determined the lensing probabilities
for isothermal spheres with finite cores;
Wallington \& Narayan (1993) investigated the influence of the
core radius on the imaging properties of elliptical lenses;
Kovner (1987a,b) looked into ``marginal" lenses;
Kassiola \& Kovner (1993) compared the lensing properties
of elliptical mass distributions with elliptical potentials.
Most of these authors studied idealized lens models in great detail. 
In order to get a good representation of, e.g., the magnification
probability of a certain elliptical lens, one needs a grid of rays
with good coverage of the source plane, in particuar of
the multiply imaged region (of order 1000$^2$ rays).

We have a very different objective here, we want to
study large scale statistical
lensing properties of the matter rather than
to model and understand individual lenses in great detail. 
Therefore our resolution is very
different from those examples mentioned above, i.e.
we have many fewer rays covering the
caustics region of our multiply imaging lenses than they have. 
This is due to the different questions we try to answer. 
Nevertheless, we like to know how well the lensing properties are 
reproduced in the high magnification regions.
Therefore we compare the lensing properties of ideal elliptical
lenses as obtained with our method to the high resolution results
obtained by others.

First we describe here briefly how we determine the magnification:
We evaluate 
the magnification in a regular grid in the source plane, as
is explained in more detail in Chapter 4.
This is done by using the differential deflection properties of the 
four rays closest to the grid position to be evaluated. 
Subsequently we interpolate two-dimensionally, to determine the
value at the desired position. 
This way we do get a moderate resolution
though certainly not a perfect  coverage of the multiply imaged 
region:
a region with size about (350 arcsec)$^2$ is covered
by about 500$^2$  resolution elements in the examples here.
However, we would like to emphasize that the resolution is 
not hardwired in this method.
It can and will be increased with the availability of faster 
computers with more memory 
or for the investigation of specific problems.

Below we present a test of the accuracy of the magnification 
distribution obtained by our method. 
For that purpose we use elliptical lenses with
different parameters (ellipticity, core radius,
velocity dispersion).
We test the accuracy of our method by using elliptical lenses 
as described by Wallington \& Narayan (1993, henceforth WN93). 
We run tests with a single lens
plane populated by an isolated elliptical lens with
various values of ellipticity, velocity dispersion and
core radius. 
The results of these tests show
that the qualitative lensing properties
(the caustic geometry,
the evolution of the caustic/critical line
structure with increasing core radius, as shown in Fig.1
of WN93) are well reproduced.

We also did a detailed quantitative comparison of the 
magnification distributions
produced by the various elliptical lenses with
the cross sections presented by  WN93 (see their Fig.3).
We determined the cross sections for lensing of 
elliptical lenses with nine different velocity dispersions ranging 
from $\sigma_v = 370$ km/sec to 1000 km/sec.
The ellipticities were chosen between 0.0 and 0.2 and
we varied the core radii from 10\% to 100\% of the
Einstein radius of a corresponding  circular symmetric
isothermal sphere.
We compared the cross sections for magnifications 
larger than 2, 4, 8, 16, 32, ... with the values obtained by
WN93.  For each ``lens" as specified by the
parameter set of velocity dispersion, ellipticity, and core radius,
we run 10 different realisations (in terms of
angle relative to the coordinate axes and position of 
center of lens).
Here is the summary of the comparison of our results with
those of WN93, for three values of the velocity
dispersion $\sigma_{v1} = 370$km/sec,
           $\sigma_{v2} = 630$km/sec,
           $\sigma_{v3} = 1000$km/sec, all with ellipticity
	   $\epsilon = 0.2$, and with various core radii.
The distributions $\sigma (> \mu)$
for magnifications  $\mu = 2$, 4 and  8 agreed within about 15\% 
or better for all three values of $\sigma_v$.
For  $\mu = 16$ the deviations were still moderate for the two
higher $\sigma_v$-values (23\% and 11\%), whereas it was worse than
that for the low velocity dispersion $\sigma_{v1}$. 
Even for  $\mu = 32$, the deviation
was only 30\% for $\sigma_{v3}$, 
but quite large for the two lower values.

This indicates that the cross sections
of gravitational lenses with velocity
dispersions at or below 1000 km/sec 
are relatively well represented in the low and intermediate
magnification regime.
At high magnifications they are not very accurately 
represented in our simulations with the given resolution.
However, the fluctuations in the cross sections are mostly
of statistical nature (the high magnification regions
are covered only by a small number of resolution
elements), as we could infer from comparisons
among the 10 realisations we had used for each parameter set. 
In the applications of our method, 
there are always {\it many} lenses in each of 
our ``lines-of sight", 
and since we will use 100 or more different lines of sight for our 
statistical analyses, the deviations relative to the WN93
values will  get (much) smaller, again,
due to the decreasing statistical noise.
Hence we do get a fair enough 
representation for the low and moderately high magnification regime.
As we are interested mainly in the global statistical properties 
rather than the local ones, this test shows that our 
approach is justified.

%
%
%
%
\section{Determination  of the Underlying Matter Distributions}

The matter evolution used in this paper is simulated by
a Particle-Mesh (PM) N-body code (Cen 1992).
But a direct, brute force calculation with such a code
is unable to cover the dynamic range needed:
a ratio of cluster-cluster separation ($100h^{-1}$Mpc)
to the lensing scale of arcseconds ($10h^{-1}$kpc)
translates to $10^{12}$ mesh points, which is far beyond 
the current generation of supercomputers.
In order to overcome this problem
a novel scheme involving convolution of a small scale
simulation box (which provides good resolution)
with a large scale simulation box (which provides good statistical
information concerning clusters)
was developed (CGOT).
We basically use their scheme of convolving two box sizes, but we add
a few essential improvements. Our scheme is described in detail 
in the Appendix.

The three dimensional matter
distribution from these calculations is given for
any cosmological time as three two-dimensional projections
onto the x-y, y-z, and x-z-planes,
with an
effective resolution of comoving 10$h^{-1}$kpc, which is necessary
in order to study the lens effects on small angular scales
of approximately 5 arcsec.
Were we to wish to study still smaller scales, hydrodynamic simulations 
(cf. e.g. Cen 1992; 
Katz, Hernquist, \& Weinberg 1993;
Navarro, Frenk \& White 1994;
Summers, Davis \& Evrard 1995;
Steinmetz \& M\"uller 1994)
would be required
since the baryonic component begins to dominate over the dark 
matter component within 10 kpc.
The essence of 
the convolution method is to note that small
regions of given over (or under) density act for most processes
as if they were in homogeneous universes of larger
(or smaller) mean density.
For details and checks of the method see the Appendix.

For a typical lensing simulation a few hundred lens planes with
size $L = 5h^{-1}Mpc$ are required.
Ideally one would like to have as many 
independent cosmological sequences as lens planes used.
This is not achievable for reasons of limited CPU time. 
But neither is it really necessary. 
The reason is that for most of the planes
only a small fraction of the matter
distribution is ``illuminated" by the bundle of light rays 
(see Figure 1).
In practice, we use ten independent evolutionary runs.
For a given lens redshift the particular run to be used for
this particular lens plane is chosen randomly, as is the projection
to be used out of the three possible ones, as is the location and
orientation of the square region that is ``illuminated" by the rays.

The computing time for such a lensing simulation
is directly proportional to the number of lens planes used. 
Most of the lens planes have densities that are below the average
density of the universe (because a few have a high overdensity), 
some are almost completely empty. 
These lens planes contribute very little to the deflection angle. 
We did some experiments in bunching together a couple of lens planes
at adjacent redshifts, 
that means adding up the matter distributions, and treating the combined
matter distribution as a single ``pseudo"-plane. 
We found that for a grouping of about 10 real lens planes 
into one pseudo lens plane, the results (see
next sections) are indistinguishable from treating all the
lens planes individually. So we concluded by running the simulations with
only 60  pseudo-planes (up to a redshift of $z_S = 3$), where each one 
represents a grouping of 10 real planes. That means, in fact that one of
our pseudo-lens planes represents a rectangular 
parallelepiped of size $5h^{-1} \times 5h^{-1} \times 50 h^{-1}$Mpc$^3$.

%
%
%
%
\section{Magnification Pattern and Magnification Distribution}

In Figure 4a (COLOR PLATE) a two-dimensional
magnification pattern is shown, representing the
magnification as a function of source position at a redshift of
$z_S = 3$ for one
particular lensing run  of our example cosmology, 
a standard CDM model with $\Omega=1$ and $\sigma_8=1.05$.  
The magnification is shown as a function 
of position in the source plane. By definition,
the average magnification is $<\mu> = 1.0$, or 
$<\Delta m> = -2.5 \log <\mu> = 0$, i.e. equal to 
the case were all the matter is smoothly distributed.
Whenever we use ``magnification" (or demagnification)
subsequently, we mean
magnification (or demagnification)
relative to a situation in which all matter is
smoothly distributed. 

The magnification as a function of position in the source plane
is shown color-coded in Figure 4a:
it increases from blue through  green and red to yellow.
The border line between green and red corresponds roughly
to the average magnification $\mu = 1.0$. 
Green regions have magnifications
slightly less than one (i.e., they are slightly de-magnified, compared 
to a universe in which all matter were smoothly distributed).
Red color indicates  magnifications  of up to one magnitude, 
yellow magnification by more than one magnitude.
Note that there is quite a large region that gets demagnified 
by a small amount, and a few relatively small spots that get quite
high magnification.

Such a  magnification pattern in the source plane is
similar to the  magnification patterns used in  microlensing
simulations at much smaller angular scales (e.g., Wambsganss, 
Paczy\'nski \& Schneider 1990). 
One difference between the magnification patterns used here and
those used in microlensing is that there one is 
interested  in the change of magnification as a function of
time (due to the relative motion of observer, lens and  source),
whereas here we only study the changes in the
magnification as a function of position.
Another difference is that here we find caustics (and multiple
images) only in rare cases, 
because we basically study
the lens properties of a more or less smooth
mass distribution; in such a situation, the surface mass density
must   be above the critical value, in order to produce caustics. 
In (quasar) microlensing, ensembles of point lenses
are studied, which always produce many (micro-)images 
and, correspondingly, many caustics.

An example of a two dimensional distribution $\mu(\vec x)$
of magnification in the
{\it image/sky plane} (cf. eq(3)) is shown in Figure 4b (COLOR PLATE).
It  shows the magnification that an image  seen at this particular
position in the sky would have.
The color assignment is the same as for Figure 4a, except that
here there are locations with black color, that means formally very
high negative values of the magnification.
The boundaries between regions with high positive and those with high
negative magnifications (transitions from yellow to black) are
easy to identify. These are the critical lines. 

This image plane magnification distribution is quite similar, 
but not identical, to the magnification distribution
in the source plane shown in Figure 4a.  
The differences are most obvious
for highly magnified
positions that correspond to multiple images. Whereas in Figure 4a
(magnification pattern in the {\it source plane}), 
the {\it total} magnification of all
images corresponding to this source positions is displayed, 
in Figure 4b the {\it sky plane} magnification of the  
different images at
the corresponding positions are shown.  From 
the distribution in Figure 4b alone, however, it is 
not obvious, which images belong to the same source position.

In Figure 5 the magnification probability distribution for the
underlying cosmological model is shown.
The top panel shows the differential distribution, the bottom panel
the integrated one.
It shows the relative frequency $f(\mu)$ of magnification $\mu$,
sampled in ten equally spaced logarithmic intervals per decade.
(In contrast to most other Figures, this one shows the result
of all 100 different realizations, rather than that of an ``example"
line-of-sight only).
In the differential distribution (top panel) there is a strong peak 
just left of magnification $\mu = 1.0$, and a power law tail 
($f(\mu) \propto \mu^{-2}$) to high
magnifications (the same power law dependence
as found in TOG for point lenses and in subsequent
work for more realistic matter distributions) until the
probability drops to zero due to the finite resolution. 
The thin solid line in Figure 5 indicates the 
magnification distribution for single image cases only, whereas the
dotted line is the one for multiple images only (the bold line
is the sum of the two).
Because of the rareness of very high magnifications, 
the high magnification end naturally is quite noisy. 

The magnification distribution in Figure 5 can be interpreted as
the ``transfer function" of the matter in the universe (i.e., for this
particular cosmological model). In other words: Any
intrinsic luminosity function of, say, quasars, will be 
folded with this magnification distribution, and what we see
as the observed quasar luminosity function is the convolution
of the intrinsic luminosity function
with this ``transfer function" of the universe (Vietri \& Ostriker 
1983).
In principle one can determine the intrinsic quasar luminosity
function for a given cosmological model, once the observed
luminosity function
and the transfer function of this cosmological model are known
(Ostriker \& Vietri 1986).
In practice this will turn out to be non-trivial, mainly
 because the observed  luminosity function of quasars is
not well known (Boyle, Shanks \&  Peterson 1988).

The effect of this transfer function is even more 
obvious when applied to 
a population of perfect ``standard candles" at high redshift:
for a given source redshift, the luminosity function (originally
a delta function in such a case)
will be broadened to exactly the shape $f(\mu)$ indicated by the 
       curve in the top part of Figure 5.
The slope of this magnification probability  function for
gravitational lensing is known
analytically in the limit of very high magnifications 
(TOG; Schneider 1987; Blandford \& Narayan 1986)
and should therefore be independent of the cosmological model.
However, in the intermediate range from small to moderate
magnifications, 
different cosmological models produce
different magnification probability distributions.

%
%
%
%
\section{Caustics, Critical Lines, 
	Multiple Images}

As described in Section 2, the positions of all rays in the image
plane and in the source plane are stored for each lensing
run. Therefore it
is straightforward to determine characteristics of the mapping 
$
\ \ {\vec y} = A {\vec x} \ \ 
$
between the positions in the image plane $\vec x$ and those in the
source plane $\vec y$ (see SEF, Chapter 5). 

Of particular interest are 
the points in the image plane \ ${\vec x}$ \ at which
the Jacobian determinant \ $\det A({\vec x}) $ \ changes its sign
(in the previous section it was pointed out that these  points
can be seen in Figure 4b as the locations where the
color changes from yellow to black).
By definition
these are the {\it critical lines}. 
In Figure 6a the critical lines
are plotted for this particular line of sight. Mapping these
critical lines onto the source plane results in the {\it caustics}.
These are shown  in Figure 6b. Comparing 
Figure 6a and 6b, it is not too difficult to find out which
caustic belongs to which critical line. 

The shapes of the critical lines and of the caustics found in these
lensing simulations are quite 
irregular, they are certainly different from the symmetric 
shapes that are obtained for simple models of lenses, e.g.
singular isothermal sphere plus external shear or
elliptic potentials (e.g., Blandford \& Kochanek 1987; 
Kormann, Schneider \& Bartelmann 1994). 
The reason is simply that the
mass distributions we use are  not at all symmetric, but rather
representing the irregular
``real" distribution of matter in the universe.

Whenever the source position is inside such a
closed caustics line,  a pair of new (usually highly magnified)
images emerges. When the source is outside, but near  a caustic,
one highly magnified image is resulting. 
In Figure 7    various image configurations are shown that 
are  produced by the caustic structure in the top right part of Figure
4a/6b. They indicate the variety of image shapes and morphologies
that can be produced by this particular caustic structure. It
ranges from a very stretched (and highly magnified) single image
at the top left through elongated and aligned double and triple 
images of similar
magnification through cases with faint central (radial) images. For 
comparison, the shape and size of a unlensed source is indicated at the
bottom right corner of the bottom right panel.

One can look at these image configurations
from two different viewpoints: if the source
is extended (e.g., a distant galaxy), then Figure 7
shows the highly extended images that could be observed: giant arcs, 
arclets, tangentially aligned images. 
On the other hand, 
if the source is an unresolved quasar, then the area of the
images reflects the magnification of the two
or more quasars that are seen. The separation between the
``centers of light" would be the distance between the
multiple images.
For comparison, 
in the lower right corner  an unlensed source is shown, which
represents magnification $\mu = 1.0$.
 
%
%
%
%
\section{Giant Arcs, Arclets,
	Tangentially Aligned Galaxies, and Shear Maps}

In this section we show the effects of gravitational lensing due
to our three dimensional ray tracing on various ensembles of sources.
In the left column of Figure 8 we show four different  fields
of sources: a regular grid of circular sources, randomly distributed
circular sources, a regular grid of elliptical sources (with fixed
ellipticity and random position angle) 
and randomly distributed elliptical sources with fixed ellipticity.
The diameter of the circular sources shown here is 
about 3 arcsec.
At the corresponding right-hand side the resulting image
configurations are shown: this is what a telescope
would see if a galaxy distribution 
at a redshift of $z_S = 1$ 
like that at the left side 
were seen
through the matter distribution in front of it.
The size of an image reflects the
magnification, since surface brightness is
conserved by gravitational lensing:
the area of the images, divided by the 
area of the corresponding source, is the magnification of 
this particular image. Since all source sizes and shapes
are identical in one panel, 
the relative areas of the images reflect the
relative magnifications.
In the panels with a regular  grid of source positions, 
it is very obvious, 
from the deviations of the chains of galaxies from straight
lines, where the largest
matter concentrations exist along the line
of sight (cf. with the map of the matter in Figure 3b).

Since all the sources have identical intrinsic sizes and shapes
in each panel (circular or constant ellipticity), each deviation
from that shape on the right hand side is lens-induced.
One can see a large variety of lens-induced image shapes:
slight ellipses from circular sources, small arcs, ``straight arcs",
giant arcs, big blobs, and more complicated  image structures.
Perhaps most obvious is the tendency of ``tangentially aligned" 
(roughly) elliptical images, whose major axes are about perpendicular
to the radius vector from the center of the mass concentration.

In Figure 9 the lensing effect of our example line of sight
is shown  for three different source redshifts:
the left columns from top to bottom show three
sets of circular sources with random positions 
at source redshifts $z_S = 1, 2, 3$.
The second column shows the respective image configurations that
one would see in the sky, if seen through the deflecting
matter between source
plane and observer. In the lowest
row, the three source populations are
superimposed, with the different redshifts indicated by three different
gray scales.

In Section 2 we described how one can obtain the 
surface mass density $\kappa$ (cf. eq. (4) )
along the line of sight, as measured by the light rays.
In Figure 4c (COLOR PLATE) we show the distribution of the effective
surface mass density $\kappa$
for our example field of view. The highest density clumps are
easily identifiable with the highest magnification regions in
Figures 4a and 4b. It is also quite obvious, how the magnification
$\mu(\vec x)$ (Figure 4b) follows the surface mass density distribution
very closely (Figure 4c) in the low magnification regime.
This indicates that it is the convergence, the matter inside the
beam, that is mainly responsible for the magnification, and not
so much the shear field from outside the beam.

For each of our artificial lines of sight, we can obtain the
distribution of the shear fields as shown in eq. (5) in Section 2. 
For our example line of sight, this  
shear distribution $\vec \gamma (\vec x)$ 
is shown for a source redshift of
$z_S = 1.0$ in Figure 4d (COLOR PLATE).
The length of the straight lines is proportional to
the absolute value of the shear, and the direction represents the
direction of the shear. It is quite obvious that the big lump
of matter in the top right part produces the strongest shear, the
lines clearly indicate the tangential alignment expected for 
galaxies that are behind such a cluster of galaxies. 
Similar shear distributions
have recently been obtained 
observationally 
by various groups who study the distribution of
shear in fields around rich clusters of galaxies 
(e.g.,  Bonnet \etal\ 1993)
or in ``empty" fields (Mould \etal\  1994).
Observationally, the shear fields are
obtained by
averaging the ellipticities  of galaxies in small fields of a few tens
of arcseconds. The idea is that this averaging should
cancel the intrinsic distribution of position angles, and as a 
net effect just show the effective shear produced by the gravitational
lens effect of the foreground matter.

%
%
%
%
\section{Summary and Outlook}

We have described a method to determine the lensing properties of a 
three dimensional, fully nonlinear
distribution of matter, obtained from cosmological
simulations. We explain the techniques we use to calculate the light
deflection by a large  number of lenses in many lens planes.
The simulations produce maps of the magnification in the source
and image plane. We can study the caustics and critical lines
for each matter realization individually, or  in a statistical way
for a large number of realizations.
Here one particular ``line of sight"  with the
matter distribution from a standard CDM model 
with $\Omega=1$ and $\sigma_8=1.05$
is used, as an example,
to illustrate diagnostic power of the applications.
It is straightforward to apply
this method to other cosmological models, 
and we plan to do so with significantly higher
resolutions in both cosmological simulations and ray-tracing method.
There are numerous possible
applications of this ray shooting technique.
Just to mention a few, which we plan to address in the future:

\noindent
1) Study of frequency, geometry, magnification ratio, 
	and separation  of multiple images. What fraction of quasars
	is multiply imaged? How many double, triple, quadruple images
	do we obtain (for given observational limits of spatial 
	resolution and dynamic range)?

\noindent
2) Study of the distribution of magnification of 
	certain classes of  objects (galaxies, quasars); 
	this can be done separately for
	single images and multiple image cases, or for both combined,
	addressing question like ``are there many
	highly magnified quasars that are single (rather than
	multiply imaged)?". Another straightforward application 
	is a quantitative study of the dispersion effect of
	lensing on standard candles at moderate
	redshifts (i.e. on supernovae of type 1).

\noindent
3) Study of correlations between single/multiple quasars  
	and strong (``visible") mass concentrations at or
	near the line of sight (which may affect the issues of
	dark lenses and/or  quasar-galaxy-associations)

\noindent
4) Study of effects of chance alignments of dense matter clumps
	along the line of sight: how accurate are the mass 
	determinations which  assume  that all matter reponsible 
	for the lensing is in a single plane?

\noindent
5) Study of the shear distribution/tangentially aligned galaxies
	around  (rich) clusters and in ``empty" fields. What is
	the average or rms-shear value in random lines of sight?

\noindent
6) Use techniques of reconstructing  mass profiles of galaxy
	clusters from weak
	lensing (Kaiser \& Squires 1993), and compare the
	reconstructed mass profile with the actual one. Are there
	any systematic effects due to foreground and background
	matter?

\noindent
7) Compare the frequency and properties of arcs and ``straight
arcs" produced by extreme lensing in clusters with that
seen in the real world.

\noindent
8) Study of the effects of three dimensional lensing on the
	cosmic microwave background. 

Most of  the points above can be investigated  for sources 
at different redshifts.
We know that the expected and computable gravitational
lensing properties of the different models,
that have been proposed for the growth of structure
in the universe, will vary greatly
from model to model.
Thus a quantitative
comparison between observed and expected lensing properties
should offer great power in
selecting
among the competing models, and ideally allow us to 
eliminate models which fail to correspond to reality.
Even at present the strongest constraints on the size
of the cosmological constant are derived from lensing statistics
(Fukugita \& Turner 1991;
Kochanek 1992;  
Maoz \& Rix 1993,
Kochanek 1995)  
In a recent application of the method
described here, we (Wambsganss \etal\ 1995) found that 
the standard CDM model predicts far more large splitting multiple 
quasars than are observed.
It will be very interesting
to see if other models such as mixed dark matter, CDM$+\Lambda$ or open
models can survive this gravitational lensing test.

\acknowledgments
It is a pleasure to thank 
Michal Jaroszy\'nski,
Avi Loeb,
Jordi Miralda-Escud\'e,
Peter Schneider, and
Ed  Turner
for many useful discussions at various stages of this project.
We also want to thank an anonymous referee for his constructive
criticism of the first version of the manuscript,
which significantly improved the quality of this paper.

\section*{Appendix: Description of the Convolution Technique} 

Here we describe the ``convolution method", which
is used to compensate the lack of sufficient spatial dynamic
range in the N-body simulations.
As stated earlier in the main body of this paper,
for gravitational lensing applications
one must have enough scale-resolving power to reach
scales which are interesting and accessible to observations
($\sim$ arcsecond). 
But it is not
possible at the same time,
in a single simulation, to reach the other 
end of the spectrum, the large
scale waves of the universe, to have a fair sample of 
the relevant cosmic systems, which are clusters of galaxies in this case.
A minimal desired dynamic range will be a few times the ratio of 
cluster separation ($\sim 50h^{-1}$Mpc) to cluster core size
($50h^{-1}$kpc), i.e., a few thousand,
which was not achievable at the time
when the simulations used in this paper were computed.

As an alternative to the brute-force approach,
we made two kinds of simulations:
one with box size $L=5h^{-1}$Mpc (Box-S hereafter) and 
resolution of $10h^{-1}$kpc, 
the other with box size $L=400h^{-1}$Mpc (Box-L hereafter) and 
resolution of $0.8h^{-1}$Mpc (both simulations use
$500^3$ cells with a PM N-body code).
In computations of both boxes (S,L) we assume periodic boundary
conditions.
That is we assume that both boxes
have the same mean density as the universe at any given redshift.
For the large box that is a good approximation since
the amplitudes of waves on scales larger than $400h^{-1}$Mpc are quite small
at $z=0$ and even smaller at the redshifts relevant for lensing.
But for the small box, this assumption is poor.
The nonlinear scale at $z=0$ is approximately $8h^{-1}$Mpc so,
in the L box we will find $5h^{-1}$Mpc subboxes at 
a variety of over and under densities.
Then using this distribution
we convolve Box-S (which has small scale resolution)
with Box-L (which has large scale power), details of which are 
described below.
We also note that, at the time this paper is being written, we are 
using new N-body techniques to make simulations which can cover
this needed range directly.
We note that the gravitational ray-tracing method presented 
in the paper is independent of the convolution method described here.

Before a step-by-step description of the convolution method,
it is useful to define a few quantities.
The actual over- or under-density  of a particular plane, 
	$R = \Sigma / < \Sigma > $,  
drawn from the density distribution $f(R)$ on the
scale of Box-S ($5^3h^{-3}$Mpc$^3$),  
which is computed from Box-L 
(containing $512,000$~ subboxes each
of which is a $5^3h^{-3}$Mpc$^3$ cube).
$R$ is the ratio of the density in the Box-S to the mean cosmic density.
The projection of the matter in a Box-S onto one of the three 
orthogonal planes makes a surface mass density screen 
	called $\Sigma_{1}(\vec x)$.
We now detail steps to do the convolution.

1) We first choose $R$ randomly from $f(R)$.

2) Then a mosaic square region, 
	$\Sigma_{2  }(\vec x)$,
consisting of $N^2$ $\Sigma_{  1}(\vec x)$ screens,
where $N=2,3,4...$, is set up.
The number of repetition screens, $N$, in each
dimension on the projection plane is determined
by $R$ such that the shrunk $\Sigma_{4  }(\vec x)$ (when $R>1$, see
step 4 below)
still has a comoving length larger than $1.5\times (5h^{-1}$Mpc)
to prevent situations of no defined densities
from happening, when rays shoot outside the original region.

3) From 
	$\Sigma_{2  }$ we obtain a new screen called,
	$\Sigma_{3  }(\vec x)$   
equal to $\Sigma_{2  }^{R^{0.07}}(\vec x)$,
where the power operation operates on each individual cell of screen
$\Sigma_{2  }$.
This step is to mimic the evolution of regions
with different densities and the value $0.07$
is empirically determined using simulations.
This step does not, to first order, change
the overall density but it
increases the contrast (for $R>1$).
Since $R^{0.07}$ is always very close to unity, this
step affect the results very mildly.

4) Next, we rescale each of the two dimensions
of the screen 
	$\Sigma_{3  }(\vec x)$ by $R^{-1/3}$.
This involves resampling the surface mass onto a smaller (or larger)
grid for $R>1$ ($R<1$), increasing (decreasing)
the surface density by $R^{2/3}$, 
resulting in a new screen, 
	$\Sigma_{4  }(\vec x)= SCALE(\Sigma_{3  } (\vec x))$.
This step is to take      account of the fact
that high density regions should have been compressed to
smaller comoving regions during their evolution.

5) The density
in each of the new pixels 
of $\Sigma_{4  }$ is multiplied by $R^{1/3}$ to take into account
   the effect that the line of sight dimension is also reduced, 
yielding 
	$\Sigma_{5  }=R^{1/3} \Sigma_{4  }$.

6) Randomly choose an origin and randomly rotate
screen $\Sigma_{5  }$ to get, by bilinear interpolation,
	a new screen
	$\Sigma_{6  }$
with the original comoving pixel size ($10h^{-1}$kpc).

7) Finally, renormalize 
	$\Sigma_{6  }$ such that
$\langle \Sigma_6\rangle = R \times \langle\Sigma_1\rangle$,
which will be the end product of this scheme.  
The renormalization factor $R$ is typically very close to unity.

In Figure 10 we illustrate this convolution. The four panels correspond
to the original surface mass density distribution $\Sigma_1$ (top left),
a part of the shrunk or expanded  distribution $\Sigma_5$ (top right),
the full $\Sigma_5$  (bottom left), and the final rotated and
normalized distribution $\Sigma_6$ (bottom right). 
In Figure 10a this is shown for  a factor $R = 1.6$ (over dense), 
in Figure 10b for $R = 0.625$ (under dense)

This convolution method is tested against a high resolution P$^3$M simulation
(Bertschinger 1995) where the boxsize is $100h^{-1}$Mpc and therefore it
contains large scale power. Of cource, this P$^3$M simulation 
does not reach the resolution of $10h^{-1}$kpc on the small scale
end, so we can only make comparisons in the overlapping region.
Figure 11  shows the surface density distributions on a cell
of area $125\times 125h^{-2}$kpc$^2$ with depth of $100h^{-1}$Mpc
using our convolution method (solid histogram) and from direct
calculation using Bertschinger's simulation (dotted histogram).
The agreement is satisfactory.

\bigskip
\bigskip

\section*{FIGURE CAPTIONS} 

\medskip
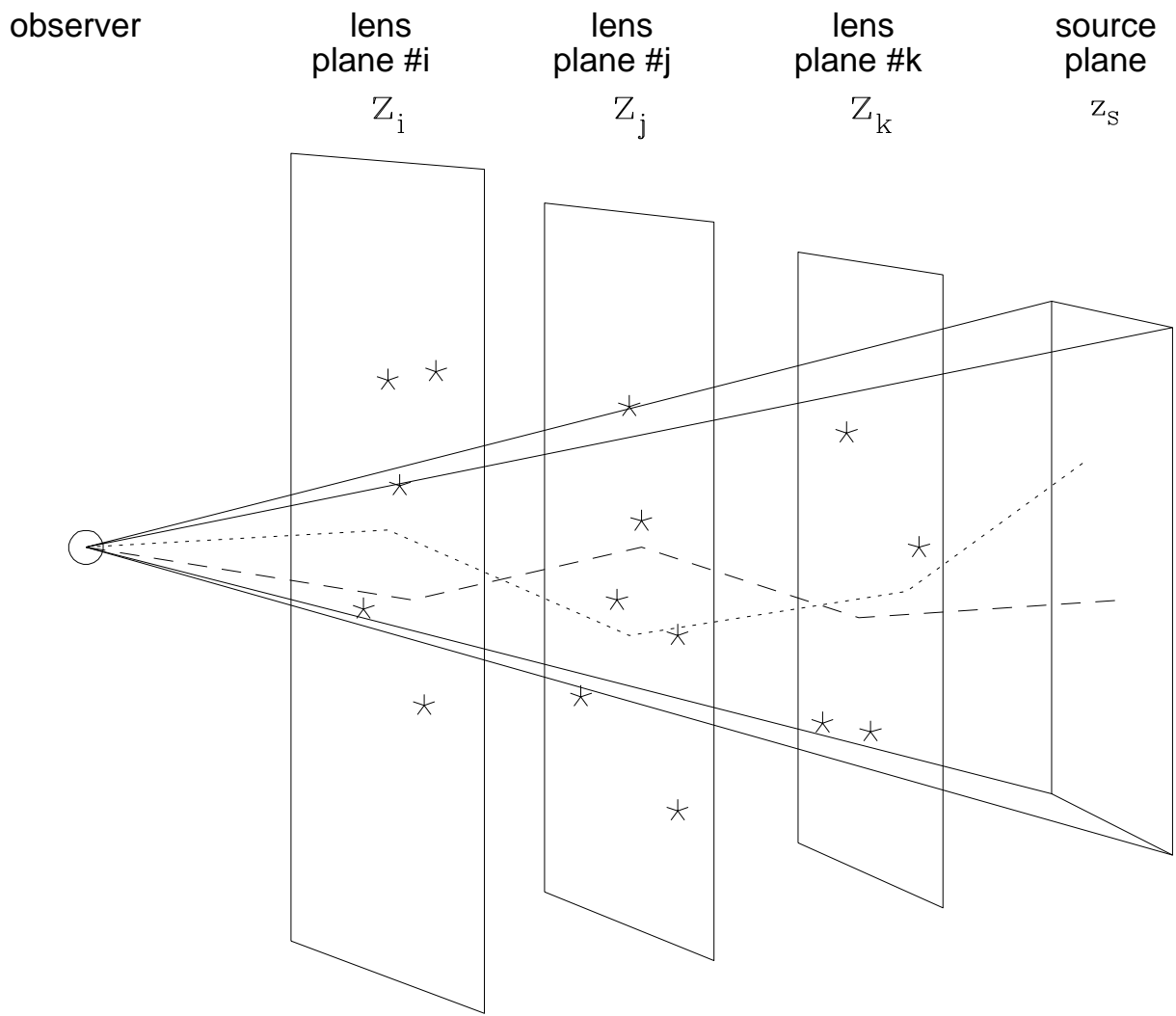
{\bf Figure 1} Schematic depiction of the bundle of light rays with
fixed opening angle traversing many planes with constant comoving
(but increasing physical) side length. 
For simplicity we show only three lens planes, in fact we use
of order 100 individual planes.
The observer is at the left hand
side, the source plane is at the right hand side.
Redshifts of the lens planes are increasing from left to right,
(i.e., $0 < z_i < z_j < z_k < z_S$).
The bundle of light rays ``illuminates" a small fraction of the lens
planes at low redshift. This fraction increases with redshift to
almost one at the highest lens redshift.
 
\medskip
{\bf Figure 2} Displayed is the effect of increasing the accuracy 
of the hierarchical
tree code for the determination of the deflection angle
by using multipoles of higher  order.
The top
(bottom) row deals with the x-(y-)component of the deflection angle. 
The difference between direct 
determination and tree-code determination is shown versus
the direct determination (in arbitrary units).  
In the leftmost column the tree code uses just the monopole term,
i.e. it assumes all the matter is at the location of the center of mass
of each cell.  The accuracy of the angle determination in this
case is about 1 \%. 
In the next columns the 
same is shown for  a tree code including quadrupole moment  (column 2)
and tree code with all moments up to order 4 (column 3). 
In the right most column all multipole moments up to order 6 
are included in the determination of the deflection angle.
The maximum deviation in the last
panel is clearly better than 0.1 \%.
All our lensing calculations are done with this accuracy.

\medskip
{\bf Figure 3} The   matter distribution in physical units 
(grams per cm$^2$) inside the beam of rays is shown. 
Dark color indicates high values of the surface mass density.
Figure 3a: 60  panels contain  the matter 
distribution of individually treated  lens planes. 
The ``first" plane is top left 
(lowest redshift), with redshift increasing to the right.
The highest redshift plane is the last one in the bottom row
at $z = 3$.
In  Figure 3b this  matter integrated along the line
of sight up to a redshift z = 3 is displayed.

\medskip
{\bf Figure 4a}  Magnification pattern  of
one  particular ``line of sight" through the universe
(the same as portrayed in Figure 3).
This is a two dimensional  map of the magnification 
as a function of position in the {\it source } plane at
$z_S = 3$.
The color indicates the {\it total}
magnification of all images for a  particular source position:
it increases from blue through green and red to yellow.
The transition  between green and red marks roughly the
average magnification.
For the strong lens in the top right part the caustics are 
identifiable as sharp yellow lines, indicating very high magnification.
The sidelength of this and all the other maps is 343 arcsec,
comparable to the HST WFPC side length of $160$ arcsec.

\medskip
{\bf Figure 4b} Magnification map in the {\it image/sky} plane. Here the
magnification of each image appearing at a certain position in the
sky is indicated. The low and intermediate
magnification regions are  very similar
to those in the map shown in Figure 4a.  
At the high magnification end there 
are obvious differences: here each image of a multiply imaged
source would 
appear at a different position. Note the transitions from bright
yellow to black: 
Here the magnification changes from very high positive values
to (formally) very high negative values. 
That means this boundary marks the ``critical
lines" which separate multiple images with positive and negative parity.

\medskip
{\bf Figure 4c} Map of effective surface mass density $\kappa$ or
convergence along the line of sight.
This distribution is quite similar to the matter integrated
along the line of sight, as shown in Figure  3b. However, here the
matter is ``weighted" by the angular diameter distances.

\medskip
{\bf Figure 4d}  Shear map  for a source plane at $z_S = 1$,
due to the foreground mass distribution in the
field of view. The length of the lines indicates the relative strength
of the shear, the angles of the lines show  the direction of 
the shear. The regions of strong shear are easily recognized as
those with the highest mass concentrations (cf. Figures 3b, 
Figures 4a/b/c).

\medskip
{\bf Figure 5}  Magnification distribution from 100 different
realizations (i.e., about $2.5 \times 10^7$ different
source positions at a redshift of $z_S = 3.0$:
bold line: all source positions; thin solid line:
singly imaged source positions only;
dashed line: multiply imaged source positions only.
The top panel displays the differential distribution
f($\mu)$, the bottom panel the integrated one
f($>\mu)$.
The typical (unlensed) source becomes slightly fainter, but there is 
a powerlaw tail of highly magnified but rare sources.

\medskip
{\bf Figure 6}
Critical lines in the image plane (Figure 6a) and caustics in the source
plane (Figure 6b) for our example line of sight
are displayed, for a source redshift of $z_S = 3.0$. 
The critical lines are the locations at which
the determinant of the Jacobian matrix disappears,
the caustics are these points,  mapped onto the source
plane (cf. with Figures 4b and 4a, respectively).

\medskip
{\bf Figure 7}  Various image configurations with large magnifications,
showing the range of possible geometrical arrangements of images.
In the bottom right part of the bottom right panel the
size and shape of the unlensed source is indicated (the relative
scale is in arcsec).

\medskip
{\bf Figure 8} Various distributions of 
sources at redshift $z_S = 1.0$
and corresponding images as seen through the three dimensional
matter distribution, i.e. weakly and strongly lensed sources.
The source configurations are always shown left, the image
configurations right. From top to bottom the source ensembles
are:
circular sources in a regular grid and randomly distributed,
and elliptical sources (random orientation) in a regular grid
and randomly placed. 
The source sizes are about 3 arcsec.
The source density is equivalent to about
40 000 galaxies per square degree. 

\medskip
{\bf Figure 9}  Similar to Figure 8, but now
source planes (left) and image planes (right) are shown 
for three different source redshifts. Starting at the top:
           $z_S = 1$,
           $z_S = 2$,
           $z_S = 3$.
	   In the bottom row the superposition of the three
	   source redshifts is shown.

\medskip
{\bf Figure 10} The panels in this figure illustrate the 
convolution technique as described in the Appendix. 
In Figure 10a
the top left panel represents the matter
distribution $\Sigma_1$ in one of the $5 h^{-1} Mpc$ cubes with average
density at redshift $z_L = 0.73$.  Here, 
this particular plane should represent a randomly determined 
overdensity of 1.6 (a larger than typical case chosen for
purpose of illustration); 
therefore, the matter is contracted in
x- and y-direction each by $^3\sqrt{1.6}$, and multiplied by
$^3\sqrt{1.6}$ (corresponding to a compression in z-direction).
The resulting matter distribution is displayed in the top right
panel (part of $\Sigma_5$, in the notation of the Appendix). 
Now, the matter distribution is extended in the ``empty"
parts of the square, taking advantage of the periodic boundary
conditions (lower left panel, full $\Sigma_5$). 
This distribution, now, is shifted
in x- and y-direction and rotated by arbitrary amounts, 
so that finally the lower right panel is obtained ($\Sigma_6$). 
The apparent 
double appearance of the two dense matter clumps is of no worry,
since the bundle of light rays with fixed angular
size cuts  out only a small square (as indicated in the final panel),
so that no repeated structures appear in the ``illuminated field".
In Figure 10b the same is shown for a (under-) density of
$ R = 1.6^{-1} = 0.625$.

\medskip
{\bf Figure 11} Comparison of surface mass density distributions
in $125h^{-1}$kpc$\times 125h^{-1}$kpc$\times 100h^{-1}$Mpc cylinders
using the convolution technique (see appendix A; solid histogram)
and a direct simulation (Bertschinger 1996; dotted histogram).
It indicates that the convolution method gives results which
are in satisfactory agreement with direct simulations, where comparisons
can be made.

\vfill\eject
\end{document}